\documentclass[conference,10pt,letterpaper]{IEEEtran}

\usepackage[utf8]{inputenc} 
\usepackage[T1]{fontenc}
\usepackage{balance}
\usepackage{cite}
\usepackage[cmex10]{amsmath}
\usepackage{booktabs}

\DeclareFontFamily{U}{mathx}{\hyphenchar\font45}
\DeclareFontShape{U}{mathx}{m}{n}{<-> mathx10}{}
\DeclareSymbolFont{mathx}{U}{mathx}{m}{n}
\DeclareMathAccent{\widebar}{0}{mathx}{"73}

\usepackage{amsmath,amsfonts,amssymb,psfrag}
\usepackage{amsthm}
\usepackage{graphicx}

\usepackage{epstopdf}
\usepackage{rotating}
\usepackage{bm}
\usepackage{dsfont}
\usepackage{color}
\usepackage{tikz}
\usetikzlibrary{plotmarks}
\usetikzlibrary{shapes,arrows,fit,calc,positioning,automata}
\usepackage{pgfplots}
\usetikzlibrary{calc}
\usetikzlibrary{decorations.markings}
\usetikzlibrary{positioning}
\pgfplotsset{compat=1.10}

\makeatletter
\newsavebox\myboxA
\newsavebox\myboxB
\newlength\mylenA

\newcommand*\xoverline[2][0.75]{%
	\sbox{\myboxA}{$\m@th#2$}%
	\setbox\myboxB\null
	\ht\myboxB=\ht\myboxA%
	\dp\myboxB=\dp\myboxA%
	\wd\myboxB=#1\wd\myboxA
	\sbox\myboxB{$\m@th\overline{\copy\myboxB}$}
	\setlength\mylenA{\the\wd\myboxA}
	\addtolength\mylenA{-\the\wd\myboxB}%
	\ifdim\wd\myboxB<\wd\myboxA%
	\rlap{\hskip 0.5\mylenA\usebox\myboxB}{\usebox\myboxA}%
	\else
	\hskip -0.5\mylenA\rlap{\usebox\myboxA}{\hskip 0.5\mylenA\usebox\myboxB}%
	\fi}
\makeatother

\usepackage{mathtools}
\usepackage{verbatim}
\usepackage[ruled,vlined,titlenumbered,linesnumbered]{algorithm2e}
\usepackage{tabularx}
\usepackage{makecell}
\usepackage{multirow}
\usepackage{dsfont}
\usepackage{array,bm}
\newcolumntype{L}[1]{>{\raggedright\let\newline\\\arraybackslash\hspace{0pt}}m{#1}}
\newcolumntype{C}[1]{>{\centering\let\newline\\\arraybackslash\hspace{0pt}}m{#1}}
\newcolumntype{R}[1]{>{\raggedleft\let\newline\\\arraybackslash\hspace{0pt}}m{#1}}
\usepackage{dsfont}

\newtheorem{theorem}{Theorem}

\newtheorem{definition}{Definition}

\newtheorem{lemma}{Lemma}

\newcommand{\plus}{\raisebox{.4\height}{\scalebox{.6}{+}}}

\IEEEoverridecommandlockouts
\begin{document}
\title{Doubly-Exponential Identification via Channels:\\ Code Constructions and Bounds}

 \author{%
   \IEEEauthorblockN{Onur G\"unl\"u\textsuperscript{1},
                     J{\"o}rg Kliewer\textsuperscript{2},
                     Rafael F. Schaefer\textsuperscript{1},
                     and Vladimir Sidorenko\textsuperscript{3} 	    
                     \thanks{This work was supported in part by the German Federal Ministry of Education and Research (BMBF) within the national initiative for ``Post Shannon Communication (NewCom)'' under the Grant 16KIS1004, U.S. National Science Foundation (NSF) under Grants 1815322 and 2107370, the European Research Council (ERC) under the European Union’s Horizon 2020 research and innovation programme (Grant Agreement No: 801434), and the Institute for Communications Engineering at TU Munich.}
                 }
   \IEEEauthorblockA{\textsuperscript{1}%
                 Chair of Communications Engineering and Security, University of Siegen, 
                  \{onur.guenlue, rafael.schaefer\}@uni-siegen.de
             }
   \IEEEauthorblockA{\textsuperscript{2}%
                     Department of Electrical and Computer Engineering, New Jersey Institute of Technology,  jkliewer@njit.edu}
    \IEEEauthorblockA{\textsuperscript{3}%
    	Institute for Communications Engineering, TU Munich,
    	vladimir.sidorenko@tum.de}
 }

\maketitle

\begin{abstract}
	Consider the identification (ID) via channels problem, where a receiver wants to decide whether the transmitted identifier is its identifier, rather than decoding the identifier. This model allows to transmit identifiers whose size scales doubly-exponentially in the blocklength, unlike common transmission (or channel) codes whose size scales exponentially. It suffices to use binary constant-weight codes (CWCs) to achieve the ID capacity. By relating the parameters of a binary CWC to the minimum distance of a code and using higher-order correlation moments, two upper bounds on the binary CWC size are proposed. These bounds are shown to be upper bounds also on the identifier sizes for ID codes constructed by using binary CWCs. We propose two code constructions based on optical orthogonal codes, which are used in optical multiple access schemes, have constant-weight codewords, and satisfy cyclic cross-correlation and auto-correlation constraints. These constructions are modified and concatenated with outer Reed-Solomon codes to propose new binary CWCs optimal for ID. Improvements to the finite-parameter performance of both our and existing code constructions are shown by using outer codes with larger minimum distance vs. blocklength ratios. We also illustrate ID performance regimes for which our ID code constructions perform significantly better than existing constructions.
\end{abstract}

\IEEEpeerreviewmaketitle
\section{Introduction}

We consider a communication problem closely related to reliable communications via a  point-to-point (P2P) channel \cite{Shannon1948,CoverandThomas}. Similar to the P2P channel model, a transmitter encodes an identifier, not known before encoding, into a codeword that is sent through a noisy channel such that a receiver observes a noisy codeword. Unlike the P2P channel model where the receiver decodes the observed noisy codeword, the receiver in the identification (ID) problem is interested in the reliable result of the \emph{binary hypothesis test} whether the transmitted identifier is the identifier of interest for him. Since the transmitted information of interest for each receiver is fixed, it is considered as an \emph{identifier} for the corresponding receiver; therefore, this hypothesis testing problem is called the \emph{identification via channels problem} \cite{AhlswedeDueck}.

One practical scenario for the ID problem is when there is a network of internet-of-things (IoT) devices, such as sensors, that are controlled by a mobile phone. Suppose we want to save energy to increase the battery life of these sensors. One way to achieve this is to insert a physical unclonable function (PUF) \cite{bizimMDPI}, which can be any digital circuit with unique outputs, into each sensor such that a uniformly distributed secret key is assigned to each device. Each secret key is an identifier for the corresponding sensor, which can be shared with the mobile phone when secure transmission is possible or by using public key cryptography. When the mobile phone intends to control a particular sensor, this sensor's identifier and the command to this sensor are encoded and broadcast through a noisy wireless channel. All sensors first apply a binary hypothesis test to decide whether they are the targeted sensor. If  this is not the case, they do not decode the command in order to save energy. Similarly, see \cite{SteinbergNerhavIDCRs} for an application of the ID problem to digital watermarking.

For a discrete memoryless channel (DMC) the ID problem is shown in \cite{AhlswedeDueck} to allow the identifier size, i.e., the number of identifiers, to be doubly-exponential in the blocklength. This is actually achievable for any channel with a non-zero transmission capacity. This is also in contrast to the P2P channel problem for which the message size is exponential in the blocklength. Reliable ID is possible for a DMC with a maximum rate being equal to its Shannon capacity \cite{AhlswedeDueck}. The main difference between the encoders for the ID and the transmission problem, in the functional sense, is that randomization increases the performance of ID codes, whereas deterministic encoders suffice for transmission. 

A source of uniformly-distributed randomness for an ID transmitter can, for example, be obtained by a PUF embodied in the transmitter; see \cite[Chapter 2]{benimdissertation}. Suppose a codeword is selected by the ID transmitter uniformly at random over the pre-determined set of codewords assigned to the identifier. There exist randomized encoding algorithms with equally sized codeword sets assigned to each identifier that achieve the \emph{ID capacity} \cite{AhlswedeDueck}. Therefore, we analyze binary constant-weight codes (CWCs), which are used to represent the equally sized codeword sets assigned to an identifier with symbol ``$1$'' and conversely codewords that cannot be chosen for a given identifier with symbol ``$0$'', respectively, as in \cite{VerduIdentification,Eswaranidentification,ChristianIdentificationEntropy}. 

An important family of binary CWCs is given by optical orthogonal codes (OOC), proposed in \cite{SalehiOOCFirst} as codes with good auto- and cross-correlation properties. OOCs are different from orthogonal (spreading) codes because OOCs consist of symbols ``$0$'' and ``$1$'', unlike orthogonal codes with symbols ``$1$'' and ``$-1$''. This property makes OOCs suitable for unipolar environments such as optical systems used for direction detection \cite{SalehiOOCFirst}, where a symbol ``$1$'' represents a detected signal and symbol ``$0$'' no signal, respectively. We modify OOCs to achieve optimality for the ID problem. We also propose a method to improve the finite-parameter performance of both our and existing ID code constructions by concatenating inner binary CWCs with suitable outer codes. Our ID code constructions significantly outperform existing constructions at low ID rates, whereas at high ID rates existing constructions perform slightly better. We next provide two finite-parameter bounds on the ID code size. We use the result from \cite[Section II-A]{AhlswedeDueck}, which states that to achieve the ID capacity, it suffices to design a binary CWC optimally for a noiseless channel and to concatenate it with a (Shannon) capacity-achieving transmission code. Thus, one can combine our proposed bounds with finite length bounds for error correction codes to obtain bounds for ID code parameters for noisy channels.

\section{Problem Formulation}\label{sec:problem_settingandcode}
\begin{figure}
	\centering
	\resizebox{0.82\linewidth}{!}{
		\begin{tikzpicture}
		\node (a) at (0,-0.5) [draw,rounded corners = 6pt, minimum width=1.5cm,minimum height=0.8cm, align=left] {\small Randomized\\ \small Encoder\\\small $Q_i(X^{n_{\text{\tiny ID}}})$};
		\node (c) at (2.3,-0.5) [draw,rounded corners = 5pt, minimum width=1.1cm,minimum height=0.6cm, align=left] {\small $P_{Y|X}$};
		\node (f) at (0,-2.05) [draw,rounded corners = 5pt, minimum width=0.7cm,minimum height=0.6cm, align=left] {\small $i\in[1\!:\!N_{\text{ID}}]$};
		\node (b) at (4.8,-0.5) [draw,rounded corners = 6pt, minimum width=2cm,minimum height=0.8cm, align=left] {\small Binary\\\small Hypothesis\\ \small Test for $i^*$};
		\draw[decoration={markings,mark=at position 1 with {\arrow[scale=1.5]{latex}}},
		postaction={decorate}, thick, shorten >=1.4pt] (a.east) -- (c.west) node [midway, above] {$x^{n_{\text{\tiny ID}}}$};
		\draw[decoration={markings,mark=at position 1 with {\arrow[scale=1.5]{latex}}},
		postaction={decorate}, thick, shorten >=1.4pt] ($(c.east)+(0.0,0)$) -- ($(b.west)+(0.0,0)$) node [midway, above] {$y^{n_{\text{\tiny ID}}}$};
		\draw[decoration={markings,mark=at position 1 with {\arrow[scale=1.5]{latex}}},
		postaction={decorate}, thick, shorten >=1.4pt] (f.north) -- (a.south) node [midway, right] {$i$};
		\node (H) at (3.8,-2.5) [rounded corners = 5pt, minimum width=1cm,minimum height=0.6cm, align=left] {$i^*\!\neq\! i$};
		\draw[decoration={markings,mark=at position 1 with {\arrow[scale=1.5]{latex}}},
		postaction={decorate}, thick, shorten >=1.4pt] (b.south) -- ($(H.north)+(0.5,0)$) node [midway, left] {$H_0$};
		\node (H1) at (6.0,-2.5) [rounded corners = 5pt, minimum width=1cm,minimum height=0.6cm, align=left] {$i^*\!=\! i$};
		\draw[decoration={markings,mark=at position 1 with {\arrow[scale=1.5]{latex}}},
		postaction={decorate}, thick, shorten >=1.4pt] (b.south) -- ($(H1.north)+(-0.5,0)$) node [midway, right] {$H_1$};
		\end{tikzpicture}
	}
\vspace*{-0.45cm}
	\caption{Identification via channels problem. The $i^*$-th receiver applies the hypothesis test for its received sequence to determine whether it is the target of the intended communication.}\label{fig:identificationproblemmodel}
	\vspace*{-0.65cm}
\end{figure}
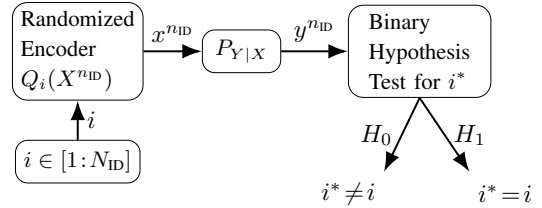

Consider $N_{\text{ID}}\geq 1$ identifiers $i\!\in\![1\!:\!N_{\text{ID}}]$, where $[1:N_{\text{ID}}]$ denotes the set $\{1,2,\ldots,N_{\text{ID}}\}$ for an $N_{\text{ID}}\in\mathbb{Z^{\plus}}$. This set represents $N_{\text{ID}}$ receivers that want to test whether they are the receiver with which the transmitter communicates. To communicate with the $i$-th receiver, the transmitter sends a sequence $X^{n_{\text{ID}}}$ whose noisy version $Y^{n_{\text{ID}}}$, associated with a DMC $P_{Y|X}$, is observed by each receiver. The $i^*$-th receiver applies a hypothesis test for its received noisy sequence to decide whether the transmitted identifier is equal to the identifier $i^*\!\in\![1\!:\!N_{\text{ID}}]$ assigned to this receiver before transmission. The null hypothesis $H_0$ for each receiver is that the transmitted identifier is not the identifier assigned to it, and the alternative hypothesis $H_1$ is that the receiver is the one with which the transmitter communicates. Fig.~\ref{fig:identificationproblemmodel} illustrates the identifier encoding procedure at the transmitter that sends $X^{n_{\text{ID}}}$ through a channel $P_{Y|X}$ and the receiver observes $Y^{n_{\text{ID}}}$ for which the hypothesis test is applied. 

There are two types of errors associated with the model shown in Fig.~\ref{fig:identificationproblemmodel}. \emph{Type-I errors} occur when the receiver mistakenly decides that it is not the receiver with which the transmitter communicates. \emph{Type-II errors} occur if the receiver mistakenly decides that it is the receiver with which the transmitter communicates. Consider a randomized encoding step that takes an identifier $i$ as input and outputs a codeword $x^{n_{\text{ID}}}\in\mathcal{X}^{n_{\text{ID}}}$ according to a probability distribution $Q_i(X^{n_{\text{ID}}}):i\rightarrow \mathcal{X}^{n_{\text{ID}}}$ for all $i\in[1:N_{\text{ID}}]$. It is shown in \cite{AhlswedeDueck} that in general a random encoder is necessary to achieve the ID capacity. Type-I and type-II errors can be characterized by defining $N_{\text{ID}}$ demapping regions $\mathcal{D}_i\subset \mathcal{Y}^{n_{\text{ID}}}$ for $i\in[1:N_{\text{ID}}]$. The randomized encoding allows to benefit from overlapping demapping regions, which is the main reason why the number of identifiers scales doubly-exponentially in the blocklength. This gain can be obtained as long as the two error probabilities can be made negligibly small \cite{Eswaranidentification}. Therefore, we define the \emph{identification via channels} problem as follows.

\begin{definition}\label{def:type1and2}
\normalfont	An $(n_{\text{ID}},N_{\text{ID}},\lambda_1, \lambda_2)$ ID code consists of $N_{\text{ID}}$ encoding probability distributions $Q_i(X^{n_{\text{ID}}})$ and demapping regions $\mathcal{D}_i\subset \mathcal{Y}^{n_{\text{ID}}}$ such that, given a DMC $P_{Y|X}$, for all $i,i^{\prime}\in[1:N_{\text{ID}}]$ and $i\neq i^{\prime}$ type-I and type-II error probabilities are upper bounded, respectively, as
	\begin{align}
		&1- \sum_{y^{n_{\text{\tiny ID}}}\in \mathcal{D}_i}\sum_{x^{n_{\text{\tiny ID}}}\in\mathcal{X}^{n_{\text{\tiny ID}}}} Q_i(x^{n_{\text{\tiny ID}}})P_{Y|X}^{n_{\text{\tiny ID}}}(y^{n_{\text{\tiny ID}}}|x^{n_{\text{\tiny ID}}})\leq \lambda_1,\\
		&\sum_{y^{n_{\text{\tiny ID}}}\in \mathcal{D}_i}\sum_{x^{n_{\text{\tiny ID}}}\in\mathcal{X}^{n_{\text{\tiny ID}}}} Q_{i^{\prime}}(x^{n_{\text{\tiny ID}}})P_{Y|X}^{n_{\text{\tiny ID}}}(y^{n_{\text{\tiny ID}}}|x^{n_{\text{\tiny ID}}})\leq \lambda_2. \label{def:idproblemdef}
	\end{align}
\end{definition}

Due to the doubly-exponential scaling of $N_{\text{ID}}$ in the blocklength $n_{\text{ID}}$, the ID rate and ID capacity are defined as follows.

\begin{definition}\label{def:IDrate}
\normalfont An ID rate $R_{\text{ID}}$ is achievable if, given any $\lambda_1,\lambda_2,\epsilon>0$, there exist some $n_{\text{ID}}\!\geq\!1$, encoding probability distributions, and demapping regions satisfying Definition~\ref{def:type1and2} and
\begin{align}
\log(\log(N_{\text{ID}})) > n_{\text{ID}}(R_{\text{ID}}-\epsilon). 
\end{align}
The \emph{ID capacity} $C_{\text{ID}}$ is the supremum over all achievable $R_{\text{ID}}$.
\end{definition}

We next state the result that the ID capacity $C_{\text{ID}}$ of a DMC $P_{Y|X}$ is equal to its Shannon capacity $C_{\text{Sh}}$.
\begin{theorem}[\hspace{1sp}\cite{AhlswedeDueck}]
	The ID capacity of a DMC $P_{Y|X}$ is 
	\begin{align}
		C_{\text{ID}} = \max_{P_X}I(X;Y) = C_{\text{Sh}}.
	\end{align}
\end{theorem}
If there is available common randomness shared between the transmitter and receiver, the ID capacity $C_{\text{ID}}$ of a DMC increases by the entropy rate of the common randomness \cite{SteinbergNerhavIDCRs}. This provides an exponential increase in the number $N_{\text{ID}}$ of identifiers with only a few bits of common randomness. Thus, the performance of any ID code construction, including our constructions below, can be significantly improved if there is a source of common randomness available such as PUFs \cite{benimMultiEntityTIFS}.

Achievability of the ID capacity $C_{\text{ID}}$ is shown in \cite[Section~II-A]{AhlswedeDueck} to be possible by using encoding probability distributions $Q_i(\cdot)$ that are uniformly distributed over equally sized support sets, which can be represented by binary CWCs \cite{VerduIdentification}. Therefore, we next define the parameters of binary CWCs along with the conditions for them to be optimal ID codes.

\begin{definition}\label{def:CWC}
	\normalfont An $(S_{\text{cw}},N_{\text{cw}},W_{\text{cw}},K_{\text{cw}})$ binary CWC consists of  $N_{\text{cw}}$ codewords $\{x_1^{S_{\text{cw}}},x_2^{S_{\text{cw}}},\ldots,x_{N_{\text{cw}}}^{S_{\text{cw}}}   \}$ of blocklength $S_{\text{cw}}$ and Hamming weight $W_{\text{cw}}$ with symbols $x_{j,s}\in\{0,1\}$  for $j=1,2,\ldots, N_{\text{cw}}$ and $s\!=\!0,1,\ldots,S_{\text{cw}}-1$ such that the maximum number of overlaps of symbols $x_{j,s}=1$ over all codeword pairs is $K_{\text{cw}}$, i.e., we have the cross-correlation
	\begin{align}
		\gamma_{j,j'}\!\triangleq\!\sum_{s=0}^{S_{\text{cw}}-1}\!x_{j,s}x_{j',s}\!\leq \!K_{\text{cw}},\!\quad \forall j,j'\!\in\![1\!:\!N_{\text{cw}}]\text{ s.t. }j\!\neq \!j'.\label{eq:definitionCWCcross}
	\end{align}
	A set of binary CWC codes is \emph{optimal for ID} if we have \cite{VerduIdentification}
	\begin{alignat}{2}
		&\frac{\log(W_{\text{cw}})}{\log (S_{\text{cw}})}\rightarrow 1&&\qquad\qquad\text{(weight factor)}\label{eq:weightfactor}\\
		&\frac{\log(\log (N_{\text{cw}}))}{\log(S_{\text{cw}})}\rightarrow 1 &&\qquad\qquad \text{(second-order rate)}\label{eq:secondorderrate}\\
		&\frac{K_{\text{cw}}}{W_{\text{cw}}}\rightarrow 0&&\qquad\qquad \text{(overlap fraction)}\label{eq:overlapfraction}. 
	\end{alignat}
\end{definition}

A closely related code family to binary ID CWCs is given by OOCs. We next define OOCs since the derivation of the bounds given below on the code size of binary ID CWCs follow similar steps as for OOCs. Our new ID code constructions also modify OOCs to improve their ID performance.

\begin{definition}
	\normalfont An $(S_{\text{ooc}}, N_{\text{ooc}}, W_{\text{ooc}},\lambda_{\text{ooc,a}},\lambda_{\text{ooc,c}})$ OOC consists of $N_{\text{ooc}}$ codewords $\{x_1^{S_{\text{ooc}}},x_2^{S_{\text{ooc}}},\ldots,x_{N_{\text{ooc}}}^{S_{\text{ooc}}}   \}$ of blocklength $S_{\text{ooc}}$ and Hamming weight $W_{\text{ooc}}$ with symbols $x_{j,s}\in\{0,1\}$ for $j=1,2,\ldots, N_{\text{ooc}}$ and $s\!=\!0,1,\ldots,S_{\text{ooc}}-1$ such that for all $\tau_{\text{a}}\!\in\! [1\!:\!(S_{\text{ooc}}\!-\!1)]$, $\tau_{\text{c}}\!\in\! [0\!:\!(S_{\text{ooc}}\!-\!1)]$, $ j,j'\!\in\! [1\!:\!N_{\text{ooc}}]$, and $j\neq j'$, we have
	\begin{align}
		&\sum_{s=0}^{S_{\text{ooc}}-1}x_{j,s}x_{j,(s+\tau_{\text{a}})}\leq \lambda_{\text{ooc},a}\qquad\quad\text{(auto-correlation)}\label{eq:OOCautocorr},\\
		&\sum_{s=0}^{S_{\text{ooc}}-1}x_{j,s}x_{j',(s+\tau_{\text{c}})}\leq \lambda_{\text{ooc},c}\qquad\quad\text{(cross-correlation)}
	\end{align}
	where $(s+\tau_{\text{a}})$ and $(s+\tau_{\text{c}})$ additions are taken modulo $S_{\text{ooc}}$.
\end{definition}

We next give bounds on the size of ID codes that can be constructed by using binary CWCs with given parameters.

\section{Upper Bounds On Binary CWC Sizes} \label{sec:UpperBounds}
We first consider the minimum distance of a binary CWC. 

\begin{lemma}\label{lem:CWCdmin}
	An $(S_{\text{cw}},N_{\text{cw}},W_{\text{cw}},K_{\text{cw}})$ binary CWC has a minimum distance $d_{\text{cw}}=2(W_{\text{cw}}-K_{\text{cw}})$.
\end{lemma}
\begin{IEEEproof}
	Since the  CWC is binary and there are at most $K_{\text{cw}}$ symbols of ``$1$'' overlapping between all codeword pairs, there are at least $(W_{\text{cw}}\!-\!K_{\text{cw}})$ symbols $x_{j,s}\!\!=\!\!1$ of each codeword overlapping with $x_{j,s}\!=\!0$ symbols of another codeword. Thus, the number of symbols that are not the same is at least $2(W_{\text{cw}}-K_{\text{cw}})$ for each codeword pair of  the  binary CWC. Since there exist two binary CW codewords that have exactly $K_{\text{cw}}$ overlapping symbols $x_{j,s}\!=\!1$, the lemma follows.
\end{IEEEproof}

\begin{theorem}\label{theo:JohnsonBasedUpperBound}
	Given a binary CWC with parameters $S_{\text{cw}}$, $W_{\text{cw}}$, and $K_{\text{cw}}$, we have 
	\begin{align}
		N_{\text{cw}} \leq\Bigg\lfloor \frac{S_{\text{cw}}}{W_{\text{cw}}}\!\Bigg\lfloor\frac{(S_{\text{cw}}\!-\!1)}{(W_{\text{cw}}\!-\!1)}\!\Bigg\lfloor\!\ldots\!\Bigg\lfloor\frac{(S_{\text{cw}}\!-\!K_{\text{cw}})}{(W_{\text{cw}}\!-\!K_{\text{cw}})}\Bigg\rfloor\!\ldots\!\Bigg\rfloor\!\Bigg\rfloor\!\Bigg\rfloor.\label{eq:JonhsonboundforCWC}
	\end{align}
\end{theorem}

\begin{IEEEproof}
	We first apply the Unrestricted Johnson Bound \cite{JohnsonBoundCWC}, \cite[Theorem~2.3.6]{JohnsonBoundbook} to a CWC with parameters $S_{\text{cw}}$, $W_{\text{cw}}$, and $d_{\text{cw}}$, which can be proved by recursively puncturing codewords. Then, by using Lemma~\ref{lem:CWCdmin}, the theorem follows.
\end{IEEEproof}

The upper bound in Theorem~\ref{theo:JohnsonBasedUpperBound} can in general be improved by treating codewords of a binary CWC as a set of sequences to bound their higher-order correlation moments. Such bounds are applied in \cite{KumarOOC} to OOCs, which compared to binary CWCs satisfy extra cyclic auto-correlation and cross-correlation constraints. Therefore, results in \cite{KumarOOC} cannot be directly used for binary CWCs. We next present a new and improved upper bound on the number of codewords of a binary CWC.

For a $d^{\prime}\!\in\![1\!:\!K_{\text{cw}}]$, define $S_{\text{cw}}^{\prime} \!=\! (S_{\text{cw}}\!-\!d^{\prime})$, $W_{\text{cw}}^{\prime} \!=\! (W_{\text{cw}}\!-\!d^{\prime})$, and $K_{\text{cw}}^{\prime} = (K_{\text{cw}}\!-\!d^{\prime})$. For $\ell\in\mathbb{Z^{\plus}}$ and $u\in[1:\ell]$, define 
\begin{align}
	C_{\ell,u} = \sum_{k=0}^u {(-1)}^k \binom{u}{k}{(u-k)}^{\ell}.
\end{align}

\begin{figure*}[t]
	\begin{align}
	N_{\text{cw}}\leq \Bigg\lfloor \frac{S_{\text{cw}}}{W_{\text{cw}}}\!\Bigg\lfloor\frac{(S_{\text{cw}}\!-\!1)}{(W_{\text{cw}}\!-\!1)}\!\Bigg\lfloor\!\ldots\!\Bigg\lfloor\frac{(S_{\text{cw}}^{\prime}+1)}{(W_{\text{cw}}^{\prime}+1)}\Bigg\lfloor\frac{{(W_{\text{cw}}^{\prime})}^\ell - {(K_{\text{cw}}^{\prime})}^\ell}{\Big(\big(\sum_{u=1}^{\ell}{\binom{W_{\text{cw}}^{\prime}}{u}}^2\,C_{\ell,u}\big)\big/\binom{S_{\text{cw}}^{\prime}}{u}\Big)-{(K_{\text{cw}}^{\prime})}^\ell}\Bigg\rfloor\Bigg\rfloor\!\ldots\!\Bigg\rfloor\!\Bigg\rfloor\!\Bigg\rfloor.
	\label{eq:corrbasedbound}
	\end{align}
	\vspace*{-0.25cm}
	\hrule
	\vspace*{-0.65cm}
\end{figure*}

\begin{theorem}\label{theo:corrbasedbound}
	Given a binary CWC with parameters $S_{\text{cw}}$, $W_{\text{cw}}$, and $K_{\text{cw}}$, we have the upper bound on $N_{\text{cw}}$ given in (\ref{eq:corrbasedbound}) on the next page for any $\ell\in\mathbb{Z^{\plus}}$ and $d^{\prime}\in[1:K_{\text{cw}}]$ such that the innermost denominator in (\ref{eq:corrbasedbound}) is positive.
\end{theorem}
\begin{IEEEproof}[Proof Sketch]
	Define the $\ell\geq 1$-th order correlation moment as
	\begin{align}
		m_{\ell}=\frac{1}{N_{\text{cw}}(N_{\text{cw}}\!-\!1)}\Bigg(\sum_{j=1}^{N_{\text{cw}}}\sum_{j'=1}^{N_{\text{cw}}}\gamma_{j,j'}^{\ell}-N_{\text{cw}}W_{\text{cw}}^{\ell}\Bigg)\label{eq:corrmoment}
	\end{align}
	where $\gamma_{j,j'}$ is as defined in (\ref{eq:definitionCWCcross}) such that $\gamma_{j,j'}\!=\!W_{\text{cw}}$ if $j\!=\!j'$. We provide a lower and an upper bound on the term $(N_{\text{cw}}\!-\!1)m_{\ell}$ by using the properties of binary CWCs so that a combination of these bounds provides the bound in (\ref{eq:corrbasedbound}). We follow entirely similar steps to the ones in \cite[Appendix A]{KumarOOC} to obtain the lower bound for binary CWCs with two main differences. First, as compared to the correlation moment defined in \cite[(A2)]{KumarOOC}, our $m_{\ell}$ definition in (\ref{eq:corrmoment}) replaces $N_{\text{cw}}S_{\text{cw}}$ terms in the factors of the denominator given in \cite[(A2)]{KumarOOC} by $N_{\text{cw}}$ since binary CWCs do not impose any cyclic correlation constraints. Second, we remove the steps  \cite[(A16)]{KumarOOC} and \cite[(A17)]{KumarOOC} that assume the cyclic auto-correlation constraints in (\ref{eq:OOCautocorr}), and we apply the Cauchy-Schwarz inequality for all cases as in \cite[(A18)]{KumarOOC} to obtain the lower bound on $(N_{\text{cw}}\!-\!1)m_{\ell}$. The upper bound on $(N_{\text{cw}}\!-\!1)m_{\ell}$ used here is $(N_{\text{cw}}\!-\!1)K_{\text{cw}}^{\ell}$. Similar steps as in \cite[Appendix B]{KumarOOC} cannot be used since they provide upper bounds for OOCs by using their cyclic correlation properties. Thus, by combining the obtained lower and upper bounds on $(N_{\text{cw}}\!-\!1)m_{\ell}$ and by applying a recursion formula for any $d'\!\in\![1\!:\!K_{\text{cw}}]$, which is applied also in the Unrestricted Johnson Bound and in \cite[Theorem~4]{KumarOOC}, the theorem follows.
\end{IEEEproof}

Combining Lemma~\ref{lem:CWCdmin} and Theorem~\ref{theo:corrbasedbound}, the bound on $N_{\text{cw}}$ in (\ref{eq:corrbasedbound}) can be written as a function of $d_{\text{cw}}$. This alternative formulation provides a lower bound on the minimum distance $d_{\text{cw}}$ of binary CWCs with given parameters $S_{\text{cw}}$, $N_{\text{cw}}$, and $W_{\text{cw}}$, which can be useful to design ID binary CWCs. 

\begin{lemma}\label{lem:NIDNcw}
	If binary CWCs are used for ID, the upper bounds in (\ref{eq:JonhsonboundforCWC}) and (\ref{eq:corrbasedbound}) on $N_{\text{cw}}$ are also upper bounds on the number $N_{\text{ID}}$ of identifiers that can be reliably identified.
\end{lemma}
\begin{IEEEproof}[Proof]
	$(S_{\text{cw}},N_{\text{cw}},W_{\text{cw}},K_{\text{cw}})$ binary CWCs concatenated with a capacity $C_{\text{Sh}}$ achieving transmission code are shown in \cite[Section II-A]{AhlswedeDueck} to be asymptotically optimal ID codes. To obtain an optimal $(n_{\text{ID}},N_{\text{ID}},\lambda_1, \lambda_2)$ ID code using this concatenation, the transmission code used for error correction should have a blocklength of $n_{\text{ID}}$ and dimension of $\log(S_{\text{cw}})$; see \cite[Section 4.1]{Eswaranidentification}. This scheme achieves $N_{\text{ID}}\!=\!N_{\text{cw}}$. This is because a given identifier $i\!\in\![1\!:\!N_{\text{ID}}]$ corresponds to a CW codeword $x_i^{S_{\text{cw}}}$ such that the transmission codewords in the uniform encoding probability distributions $Q_i(x^{n_{\text{\tiny ID}}})$ are represented by symbols $x_{j,s}\!=\!1$ of the CW codeword $x_i^{S_{\text{cw}}}$, i.e., every  $x_i^{S_{\text{cw}}}$ can choose $W_{\text{CW}}$ transmission codewords.
\end{IEEEproof}

\section{Modified OOC Constructions for ID}\label{sec:OOCConstructions}
There are only a few constructive methods proposed for the ID via channels problem. In \cite{VerduIdentification,KoetterIdentification,Eswaranidentification,ChristianIdentificationEntropy} algebraic codes such as inner pulse position modulation (PPM) codes, which are binary CWCs with $W_{\text{cw}}\!=\!1$ and $K_{\text{cw}}\!=\!0$, concatenated with two outer codes are constructed to obtain binary CWCs optimal for ID. Similarly, in \cite{HashBasedIDCodes} $\epsilon$-almost strongly universal hash functions are concatenated with an outer code. These constructions concatenate a set of inner binary CWCs with one or more outer codes such that the constraints in (\ref{eq:weightfactor})-(\ref{eq:overlapfraction}) are satisfied for the concatenated set of binary CWCs; see the following lemma for the parameters of such a concatenation.

\begin{lemma}[\hspace{1sp}\cite{Eswaranidentification}]\label{lem:concatenation}
	Consider the concatenation of an inner  $(S_{\text{icw}},N_{\text{icw}},W_{\text{icw}},K_{\text{icw}})$ binary CWC with an outer error correction code with blocklength $n_{\text{o}}$, code dimension $k_{\text{o}}$, minimum distance $d_{\text{o}}$, i.e., an $(n_{\text{o}}, k_{\text{o}},d_{\text{o}})$ code. The resulting code is an $(S_{\text{icw}}n_{\text{o}},\; N_{\text{icw}}^{k_{\text{o}}},\; W_{\text{icw}}n_{\text{o}},\; W_{\text{icw}}(n_{\text{o}}-d_{\text{o}})\!+\!K_{\text{icw}}n_{\text{o}})$ binary CWC.
\end{lemma}

Lemma~\ref{lem:concatenation} suggests that to achieve a small overlap fraction, defined in (\ref{eq:overlapfraction}), the outer error correction code should have a large minimum distance vs. blocklength ratio $d_{\text{o}}/n_{\text{o}}$, whose maximum $(n_{\text{o}}\!-\!k_{\text{o}}\!+\!1)/n_{\text{o}}$ is obtained by maximum distance separable (MDS) codes. In \cite{Eswaranidentification}, $[q_{\text{o}}\!-\!1,k_{\text{o}}]$ Reed-Solomon (RS) codes over GF$(q_{\text{o}})$, which are $(q_{\text{o}}\!-\!1, k_{\text{o}}, q_{\text{o}}\!-\!k_{\text{o}})$ error correction codes with $k_{\text{o}}\!<\!q_{\text{o}}\!-\!1$ and a prime power $q_{\text{o}}$, are used  as outer codes. In \cite{VerduIdentification,ChristianIdentificationEntropy,HashBasedIDCodes}, $[q_{\text{o}},k_{\text{o}}]$ extended RS codes with parameters $(q_{\text{o}}, k_{\text{o}}, q_{\text{o}}-k_{\text{o}}+1)$ are used as outer codes, providing a larger minimum distance vs. blocklength ratio than RS codes because we have that $(q_{\text{o}}\!-\!k_{\text{o}}\!+\!1)/q_{\text{o}}>(q_{\text{o}}\!-\!k_{\text{o}})/(q_{\text{o}}\!-\!~1)$. This extension decreases the overlap fraction value. To further decrease the overlap fraction for the same field size $q_{\text{o}}$, we propose to use $[q_{\text{o}}\!+\!1,k_{\text{o}}]$ \emph{doubly-extended RS codes} that are MDS with parameters $(q_{\text{o}}\!+\!1, k_{\text{o}}, q_{\text{o}}\!-\!k_{\text{o}}\!+\!2)$ as outer codes.

We next propose modified OOC constructions adapted to the ID via channels problem as new inner binary CWCs such that their concatenations with outer (doubly-extended) RS codes are optimal. A requirement to use Lemma~\ref{lem:concatenation} for outer (doubly-extended) RS codes is to set $q_{\text{o}}\!=\!N_{\text{icw}}$ such that each symbol of the outer code can be represented as a different codeword of the inner code \cite{VerduIdentification}. Therefore, we propose modified OOC constructions with code sizes $N_{\text{icw}}$ that are prime powers.

\textbf{Construction 1}: Prime sequences are proposed in \cite{PrimeSeq1,PrimeSeq2} as a $(p^2,p,p,p\!-\!1,2)$ OOC, where $p$ is a prime. A prime sequence is generated by multiplying in modulo-$p$ all field elements of GF$(p)$ with one of the field elements, where we map each field element to an integer in the range $[0\!\!\!:\!\!\!p\!\!-\!\!1]$. For instance, prime sequences for $p~\!\!\!=~\!\!\!\!5$ are $\{(00000),(01234), (02413), (03142),(04321)\}$. Each symbol is then mapped to an index in a binary sequence of length $p$ such that at the corresponding index there is the symbol ``$1$'' and the other indices contain symbol ``$0$''. This symbol-to-binary-sequence mapping is called \emph{one-hot encoding}. For instance, the prime sequence $(01234)$ is mapped to the binary sequence $(10000\;01000\;00100\;00010\;00001)$. The number of pairwise overlaps of symbols $x_{j,s}\!=\!1$ over the binary representations of prime sequences is $K_{\text{icw}}=1$ due to the first symbol being symbol ``$0$'', common in all prime sequences. We remove this ``$0$'' (i.e., for $p=5$, we have sequences $\{(0000),(1234), (2413), (3142),(4321)\}$) to obtain binary representations of modified prime sequences that constitute a $(p^2\!-\!p,p,p\!-\!1,0)$ binary CWC, where $p$ is prime. 

If modified prime sequences are doubly concatenated with an outer $[p\!-\!1,k_\text{o}]$ RS code over GF$(p)$ and again with another outer $[p^{k_{\text{o}}}\!-\!1,k_{\text{oo}}]$ RS code over GF$(p^{k_{\text{o}}})$ (\emph{the second outer RS code}), we obtain a binary CWC with 
\begin{align}
	&S_{\text{cw}} = p{(p-1)}^2(p^{k_{\text{o}}}-1), \qquad\qquad N_{\text{cw}} = p^{k_{\text{o}}k_{\text{oo}}},\label{eq:Constr1Scw}\\
	& W_{\text{cw}} = {(p-1)}^2(p^{k_{\text{o}}}-1),\\
	& K_{\text{cw}} = {(p-1)}^2(k_{\text{00}}-1)+(p-1)(k_{\text{o}}-1)(p^{k_{\text{o}}}-1)\label{eq:Constr1Kcw}
\end{align}
which follows from Lemma~\ref{lem:concatenation}. It is straightforward to show that the binary CWCs constructed from modified prime sequences are optimal for ID if $\log(k_{\text{oo}})\!\rightarrow\!\infty$, $\log(k_{\text{oo}})/k_{\text{o}}\!\rightarrow\! 1$, $k_{\text{o}}/p\!\rightarrow\!0$, and $k_{\text{oo}}/p^{k_{\text{o}}}\!\rightarrow\! 0$. The last two conditions require the (first-order) code rates of outer codes to be asymptotically zero although the construction is optimal for ID, i.e., the second-order rate is optimal. Furthermore, the second outer RS code we use is more general than the one used in \cite{VerduIdentification,ChristianIdentificationEntropy, Eswaranidentification,HashBasedIDCodes}, where the code dimension is enforced to be $k_{\text{oo}}\!=\!p^t$ for some $t\!\in\![1\!:\!k_{\text{o}}\!-\!1]$. Thus, our optimality conditions for ID are more general than the conditions in \cite[Proposition 3]{VerduIdentification}. 

If the outer RS codes are replaced with corresponding doubly-extended RS codes, then we obtain a binary CWC with parameters in  (\ref{eq:Constr1Scw})-(\ref{eq:Constr1Kcw}) after replacing the ${(p\!-\!1)}^2$ terms with $(p^2\!-\!1)$ and $(p^{k_{\text{o}}}\!-\!1)$ terms with $(p^{k_{\text{o}}}\!+\!1)$, respectively. The asymptotic optimality conditions for ID are the same for constructions with two outer RS codes and doubly-extended RS codes.
However, using doubly-extended RS codes decreases the overlap fraction as compared to RS codes. Therefore, the type-II error probability $\lambda_2$ of the ID code, which can be obtained by concatenating the binary CWC with a capacity-achieving transmission code, also decreases by using outer doubly-extended RS codes. This is because $\lambda_2$ is shown in \cite[Proposition 1]{KoetterIdentification} to be equal to the sum of overlap fraction of the binary CWC and the block error probability of the capacity-achieving transmission code. This result suggests that binary CWC constructions that have outer codes with large minimum distance vs. blocklength ratio $d_{\text{o}}/n_{\text{o}}$ should be used to decrease $\lambda_2$ of the ID code. Furthermore, doubly-extended RS codes can be obtained by adding two parity check symbols to RS codes, which has only small extra encoding complexity. 

\textbf{Construction 2}: The following sequences are proposed in \cite{KumarOOC} as $(p^{2m}\!-\!1,p^m-2,p^m\!+\!1,2,2)$ OOCs, where $p$ is a prime and $m\in\mathbb{Z^{\plus}}$. Let $\alpha$ be a primitive element of GF$(p^{2m})$ and consider $p^m\!-\!2$ sets with elements $x$ satisfying
\begin{align}
	{(x-1)}^{p^m+1}=\alpha ^{i(p^m+1)}\label{eq:construction2equality}
\end{align}
for $i\!\in\![1\!:\!p^m\!-\!2]$, where we then map each nonzero $x$ to an integer equal to the exponent with respect to $\alpha$, i.e., we calculate the integer $\log_{\alpha}(x)$, in modulo-$(p^{2m}\!-\!1)$. We obtain $p^m\!-\!2$ sets each containing $p^m\!+\!1$ integers in the range $[1:p^{2m}\!-\!1]$ that correspond to the indices at which a binary CW codeword of blocklength $p^{2m}\!-\!1$ has the symbol ``$1$''. Since the field elements satisfying (\ref{eq:construction2equality}) are different for different $i$, this construction provides $(p^{2m}\!-\!1, p^m\!-\!2, p^m\!+\!1,0)$ binary CWCs, where $p$ is prime and $m\in\mathbb{Z^{\plus}}$.

We now can concatenate these binary CWCs with outer codes such as RS codes to obtain optimal parameters for ID. However, unlike in Construction 1, $N_{\text{icw}}\!=\!q_{\text{o}}\!=\!p^m\!-\!2$ is not a prime power for all $(p,m)$ pairs. For instance, $(p,m)\!=\!(2,\forall m\!\geq \!3),(3,7),(3,8),(5,3),(11,2),(23,3)$ do not result in prime power values $N_{\text{icw}}$, whereas various pairs such as $(p,m)\!=\!(2,2),(3,m\!\!\in\![2\!:\!6]),(3,9),(7,2),(13,2),(19,2)$ do. Thus, if (doubly-extended) RS codes are used as outer codes, it is necessary to check the prime power condition since there may not exist a general condition in the literature to obtain prime powers of the form $p^m\!-\!2$ from a prime $p$ and $m\!\in\!\mathbb{Z^{\plus}}$. One can alternatively decrease the size $N_{\text{icw}}$ of this binary CWC to the maximum prime power $p^{\prime}$ such that $p^{\prime}\leq p^m\!-\!2$.

If binary sequences obtained from the solution of (\ref{eq:construction2equality}) are doubly concatenated with an outer $[p^m\!-\!3,k_\text{o}]$ RS code over GF$(p^m\!-\!2)$ and again with another outer $[{(p^m\!-\!2)}^{k_{\text{o}}}\!-\!1,k_{\text{oo}}]$ RS code over GF$({(p^m\!-\!2)}^{k_{\text{o}}})$, we obtain binary CWCs that are optimal for ID if the same four conditions given above for the optimality of Construction 1 are satisfied here as well. Furthermore, the type-II error probability $\lambda_2$ of the ID codes constructed from these binary CWCs can be decreased by using outer codes with larger minimum distance vs. blocklength ratios $d_{\text{o}}/n_{\text{o}}$ than RS codes, as discussed for Construction~1. 

\section{ID Code Comparisons}
ID codes that consist of $(S_{\text{cw}},N_{\text{cw}},W_{\text{cw}},K_{\text{cw}})$ binary CWCs and a capacity $C_{\text{Sh}}$ achieving transmission code are asymptotically optimal ID codes \cite[Section II-A]{AhlswedeDueck} with $N_{\text{ID}}=N_{\text{cw}}$, as discussed in the proof of Lemma~\ref{lem:NIDNcw}. Thus, we consider noiseless channels $P_{Y|X}(y|x)\!=\!\mathds{1}\{x=y\}$, where $\mathds{1}\{\cdot\}$ is the indicator function. For these channels, the capacity-achieving transmission code has a code rate of $C_{\text{Sh}}\!=\!1$ symbol/channel-use, so we have $n_{\text{ID}}\!=\!\log(S_{\text{cw}})$. Furthermore, the type-I error probability is zero, i.e., $\lambda_1\!=\!0$, and the type-II error probability is upper bounded by the overlap fraction of the binary CWC, i.e., $\lambda_2\!\leq\! K_{\text{cw}}/W_{\text{cw}}$. Define the type-I and II error exponents as $\text{E}_1\!=\!-\log(\lambda_1)/n_{\text{ID}}$ and $\text{E}_2\!=\!-\log(\lambda_2)/n_{\text{ID}}$, respectively. 

\begin{theorem}[\hspace{1sp}\cite{VerduIdentification,AhlswedeDueck}]\label{theo:tighterrorexpbound}
	If there exists an $(n_{\text{ID}},N_{\text{ID}},\lambda_1, \lambda_2)$ ID code that achieves the triple $(R_{\text{ID}},\text{E}_1,\text{E}_2)$ with $\text{E}_1>0$ for a DMC $P_{Y|X}$ with channel capacity $C_{\text{Sh}}$ , then $R_{\text{ID}}+2\text{E}_2\leq C_{\text{Sh}}$. This bound is tight for noiseless channels.
\end{theorem}

\begin{figure}[t]
	\centering
	\vspace*{-3.2cm}
	\includegraphics[width=0.513\textwidth, height=0.471\textheight,keepaspectratio]{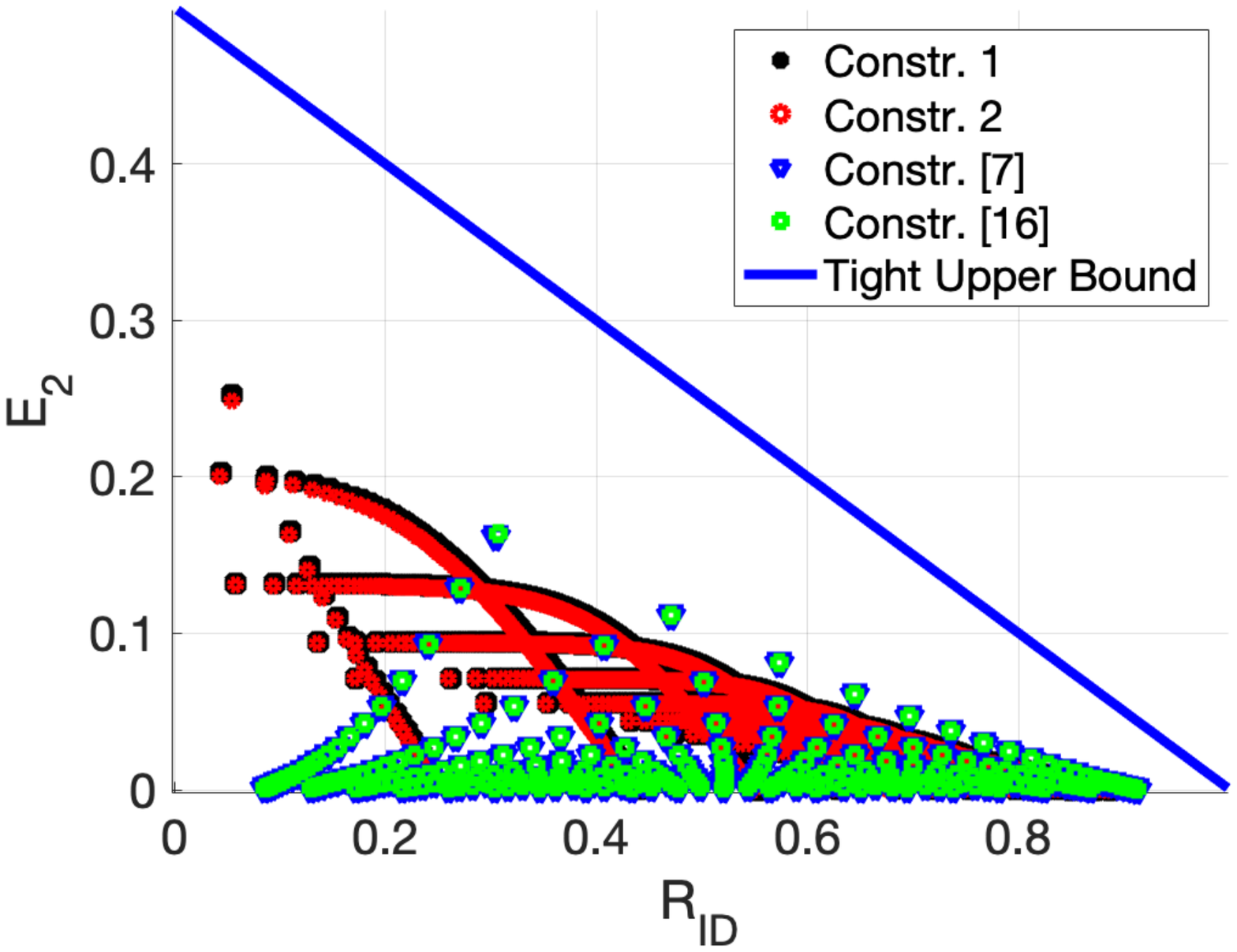}
	\vspace*{-3.2cm}
	\caption{Achieved (ID rate, type-II error exponent) tuples and the tight upper bound for a noiseless channel with $p_{\text{Constr.}1}\!=\!p_{\text{Constr.}2}^m\!-\!2=23$.} 
	\label{fig:comparisons}
	\vspace*{-0.75cm}
\end{figure}

We compare Constructions 1 and 2 with the best existing ID constructions to illustrate the achieved $(R_{\text{ID}},\text{E}_2)$ tuples for a noiseless channel. As benchmark schemes, we consider the CWC construction in \cite{VerduIdentification}, where a PPM code is concatenated with two outer extended RS codes, and in \cite{HashBasedIDCodes}, where $\epsilon$-almost strongly universal hash functions are concatenated with an outer extended RS code, respectively. The choice of the finite field used for Constructions 1 and 2 affects the encoding complexity. We therefore choose the parameters $p_{\text{Constr.}1}\!=\!p_{\text{Constr.}2}^m\!-\!2$ to have the same finite fields for both constructions, where $p_{\text{Constr.}1}$ is the parameter $p$ for Construction 1 and $p_{\text{Constr.}2}$ is the parameter $p$ for Construction 2, respectively. We assign $p_{\text{Constr.}2}\!=\!5$ and $m\!=\!2$ for Construction 2, and $p_{\text{Constr.}1}\!=\!23$ as the parameter $p$ for both Construction 1 and the constructions in \cite{VerduIdentification,HashBasedIDCodes}. Fig.~\ref{fig:comparisons} depicts the $(R_{\text{ID}},\text{E}_2)$ tuples achieved by these four constructions in addition to the tight upper bound given in Theorem~\ref{theo:tighterrorexpbound}; see \cite{BurnashevYamamotoID} for its extensions to ID of multiple identifiers. We remark that all four constructions achieve the upper bound given in  Theorem~\ref{theo:tighterrorexpbound} asymptotically.

Fig.~\ref{fig:comparisons} illustrates that Constructions 1 and 2 achieve rate tuples that are close, and Construction 1 achieves slightly larger $R_{\text{ID}}$ and $\text{E}_2$ values than Construction 2. Tuples achieved by Constructions 1 and 2 follow a similar pattern, whereas code constructions in \cite{VerduIdentification} and \cite{HashBasedIDCodes} follow a pattern that is different from the patterns of Constructions 1 and 2. Furthermore, at low ID rates $R_{\text{ID}}$ Constructions 1 and 2 achieve significantly larger type-II error exponents $\text{E}_2$ than being achieved by existing constructions, but at high ID rates the constructions in \cite{VerduIdentification} and \cite{HashBasedIDCodes} can achieve slightly larger type-II error exponents. Thus, the choice of the ID code construction should depend on the required ID rate and the allowed encoding complexity. 

\IEEEtriggeratref{11}
\bibliographystyle{IEEEtran}
\bibliography{references}

\end{document}